\documentclass[prb,aps,twocolumn,superscriptaddress,showpacs]{revtex4-1}

\usepackage{graphicx}
\usepackage{epstopdf}

\begin{document}

\title{Fluctuating defects in the incipient relaxor K$_{1-x}$Li$_x$TaO$_3$ ($x=0.02$)}

\author{C. Stock}
\affiliation{School of Physics and Astronomy, University of Edinburgh, Edinburgh EH9 3JZ, United Kingdom}
\author{P.M. Gehring}
\affiliation{NIST Center for Neutron Research, National Institute of Standards and Technology, Gaithersburg, Maryland 20899-6100, USA}
\author{G. Xu}
\affiliation{Condensed Matter Physics and Materials Science Department, Brookhaven National Laboratory, Upton, New York 11973, USA}
\author{D. Lamago}
\affiliation{Karlsruher Institut fur Techologie, Institut fur Festkorperphysik, P.O. Box 3640, D-76021 Karlsruhe, Germany}
\affiliation{Laboratoire Leon Brillouin, CEA Saclay F-91191 Gif-sur-Yvette, France}
\author{D. Reznik}
\affiliation{Karlsruher Institut fur Techologie, Institut fur Festkorperphysik, P.O. Box 3640, D-76021 Karlsruhe, Germany}
\affiliation{Laboratoire Leon Brillouin, CEA Saclay F-91191 Gif-sur-Yvette, France}
\affiliation{Department of Physics, University of Colorado, Boulder, Colorado  80309-0390, USA}
\author{M. Russina}
\affiliation{Helmholtz Zentrum Berlin fr Materialien and Energie GmbH, 14109 Berlin, Germany}
\author{J. Wen}
\affiliation{Condensed Matter Physics and Materials Science Department, Brookhaven National Laboratory, Upton, New York 11973, USA}
\author{L.A. Boatner}
\affiliation{Center for Radiation Detection Materials and Systems, Oak Ridge National Laboratory, Oak Ridge, Tennessee 37831-6032, USA}
\date{\today}

\begin{abstract}

We report neutron scattering measurements of the structural correlations associated with the apparent relaxor transition in K$_{1-x}$Li$_x$TaO$_3$ for $x=0.02$ (KLT(0.02)).  This compound displays a broad and frequency-dependent peak in the dielectric permittivity, which is the accepted hallmark of all relaxors.  However, no evidence of elastic diffuse scattering or any soft mode anomaly is observed in KLT(0.02) [J.\ Wen \emph{et al.}, Phys.\ Rev.\ B {\bf{78}}, 144202 (2008)], a situation that diverges from that in other relaxors such as PbMg$_{1/3}$Nb$_{2/3}$O$_3$.  We resolve this dichotomy by showing that the structural correlations associated with the transition in KLT(0.02) are purely dynamic at all temperatures, having a timescale on the order of $\sim$\,THz.  These fluctuations are overdamped, non-propagating, and spatially uncorrelated.  Identical measurements made on pure KTaO$_3$ show that they are absent (within experimental error) in the undoped parent material.  They exhibit a temperature dependence that correlates well with the dielectric response, which suggests that they are associated with local ferroelectric regions induced by the Li$^+$ doping.  The ferroelectric transition that is induced by the introduction of Li$^+$ cations is therefore characterized by quasistatic fluctuations, which represents a stark contrast to the soft harmonic-mode-driven transition observed in conventional perovskite ferroelectrics like PbTiO$_3$.  The dynamic, glass-like, structural correlations in KLT(0.02) are much faster than those measured in random-field-based lead-based relaxors, which exhibit a frequency scale of order of $\sim$\,GHz and are comparatively better correlated spatially.  Our results support the view that random fields give rise to the relaxor phenomena, and that the glass-like dynamics observed here characterize a nascent response.

\end{abstract}


\maketitle

\section{Introduction and comparison with lead-based relaxors}

Relaxor ferroelectrics are technologically important because they exhibit enormous room-temperature piezoelectric coefficients and sizable dielectric constants with relatively little hysteresis.~\cite{Park97:82,Ye98:81}  These materials are typically characterized by a dielectric permittivity that peaks broadly with temperature and displays a frequency-dependent amplitude and position spanning at least 14 decades in frequency.~\cite{Bovtun26:06}  This behavior differs markedly from the sharp, frequency-independent dielectric response observed in conventional ferroelectrics such as PbTiO$_3$.~\cite{Shirane50:6,Shirane70:2,Hlinka06:73,Kempa06:79} The most studied relaxor ferroelectrics are the lead-based compounds having the general formula Pb$B$O$_3$, for which the $B$-site is occupied by one of two or more heterovalent cations.  Two such relaxors are Pb(Mg$_{1/3}$Nb$_{2/3}$)O$_3$ (PMN) and Pb(Zn$_{1/3}$Nb$_{2/3}$)O$_3$ (PZN), both of which share the PbTiO$_3$ cubic perovskite structure.~\cite{Cowley11:60,Gehring13:2}  Understanding the origin of the unusual dielectric properties of these materials is a topic of considerable current interest.  To help clarify the origin of the ``relaxor transition" in the lead-based compounds, and to guide the development of new materials with similar dielectric properties, it is instructive to compare the physical properties of PMN and PZN to lead-free systems that display a similar frequency-dependent dielectric response.  To this end we present a neutron inelastic scattering study of the relaxor K$_{1-x}$Li$_x$TaO$_3$ (KLT) for $x=0.02$ (KLT(0.02)) and compare our results with those from prior studies of PMN-based and PZN-based relaxors as well as the undoped (non-relaxor) parent compound KTaO$_3$.

Neutron inelastic scattering methods have played a historically significant role in shaping our understanding of ferroelectric materials.  Studies of the lattice dynamics, and specifically of the soft, zone-center transverse optic (TO) mode, have shown that a direct link can be made between the neutron inelastic scattering cross section and the bulk dielectric response.~\cite{Lyddane41:59}  Early neutron scattering studies of several lead-based relaxors, ~\cite{Gehring01:87,Waki02:65} which were later corroborated with optical measurements,~\cite{Kamba05:17} discovered a strongly-damped, soft, TO phonon branch that was associated with the formation of local polar nanoregions (PNR) (elastic diffuse scattering).  The unusual damping of the soft TO phonon branch, which was termed the ``waterfall" effect, was initially considered to be a signature of relaxor behavior.  However subsequent studies found similarly damped TO modes in heavily-doped 40\%Pb(Mg$_{1/3}$Nb$_{2/3}$)O$_3$-60\%PbTiO$_3$ (PMN-60\%PT)~\cite{Stock06:73} and Pb(Zr$_{1-x}$Ti$_x$)O$_3$.~\cite{Phelan14:111}  Neither of these materials is a relaxor, and both show a sharp, well-defined ferroelectric transition similar to that in pure PbTiO$_3$.  Therefore the waterfall effect, though somewhat unusual for perovskites, cannot be regarded as a dynamic signature of the relaxor phase.

Neither neutron nor x-ray scattering studies have found any evidence of a truly long-range ordered, structural phase transition in PMN, as would be manifested by the splitting of a nuclear Bragg peak.  There is, however, a striking lack of consensus regarding PZN.  Several groups have reported a structural transition in PZN from cubic to rhombohedral symmetry.~\cite{Fujishiro00:69,Lebon02:14,Kisi05:17}  But other groups, using high-energy x-ray scattering techniques, have presented compelling evidence that this transition is confined to the near-surface region or ``skin" of the crystal, and that the actual shape of the PZN unit cell within the interior of the crystal remains cubic at all temperatures.~\cite{Xu03:67,Xu04:84}  The conclusion of a skin effect in PZN remains controversial.~\cite{Xu06:79,Kisi05:17}  However similar observations have also been reported in pure PMN,~\cite{Conlon04:70} as well as various compositions of PMN and PZN doped with PbTiO$_3$.~\cite{Gehring04:16,Xu03:68,Xu04:84}  Most recently a skin effect has also been observed in the lead-free perovskite relaxor Na$_{1/2}$Bi$_{1/2}$TiO$_3$ (NBT).~\cite{Ge13:88} In all cases, the skin effect is observed in systems that possess strong, quenched, random electric fields.

Despite the anomalous soft mode behavior and the controversial structural properties, neutron and x-ray scattering experiments on PMN and PZN have revealed two unique features that do appear to be experimentally linked to the relaxor state.  These features provide the motivation for our study of KLT(0.02).  First, both PMN and PZN exhibit strong, temperature-dependent, elastic (i.\ e.\ static) diffuse scattering that is polar in nature and highly anisotropic.  This has been demonstrated by x-ray scattering measurements in zero and non-zero applied electric field,~\cite{You97:79,Stock07:76} neutron scattering studies,~\cite{Vak95:37,Vak98:40,Nab99:11,Hirota02:65,Hiraka04:70,Xu04:70,Xu04:69,Pasciak12:85} and studies of the low-energy transverse acoustic (TA) phonons.~\cite{Stock12:86,Koo02:65,Nab99:11}  The diffuse scattering is not purely static at all temperatures and in fact displays a slow relaxational character on a GHz timescale that was identified using neutron spin-echo~\cite{NSE:book} and backscattering~\cite{BS:1998,BS:2003} techniques.~\cite{Stock10:81,Xu10:82,Xu12:86}  In particular, the neutron spin-echo experiment by Stock {\em et al}.\ revealed a dynamic component to the diffuse scattering in PMN that they related to the peak response in the dielectric permittivity.~\cite{Stock12:86}  A summary of how the diffuse scattering may be a general property common to all relaxors, and not just to the lead-based systems, is given in Ref.~\onlinecite{Ge13:88}.  The second unique feature of relaxors concerns the presence of so-called ``columns" of inelastic scattering, broadly distributed in energy, that are located at the $M$-point ($\vec{Q}=({1\over2},{1\over2},0)$) and $R$-point ($\vec{Q}=({1\over2},{1\over2},{1\over2})$) zone boundaries.~\cite{Swainson09:79}  A similar zone-boundary soft mode was observed at the $M$-point with x-ray inelastic scattering methods in the disordered perovskite PZT.~\cite{Hlinka11:83}  The goal of our study of KLT(0.02) is to determine how universal these features are to the relaxor phase.

\begin{figure}[t]
\includegraphics[width=9cm] {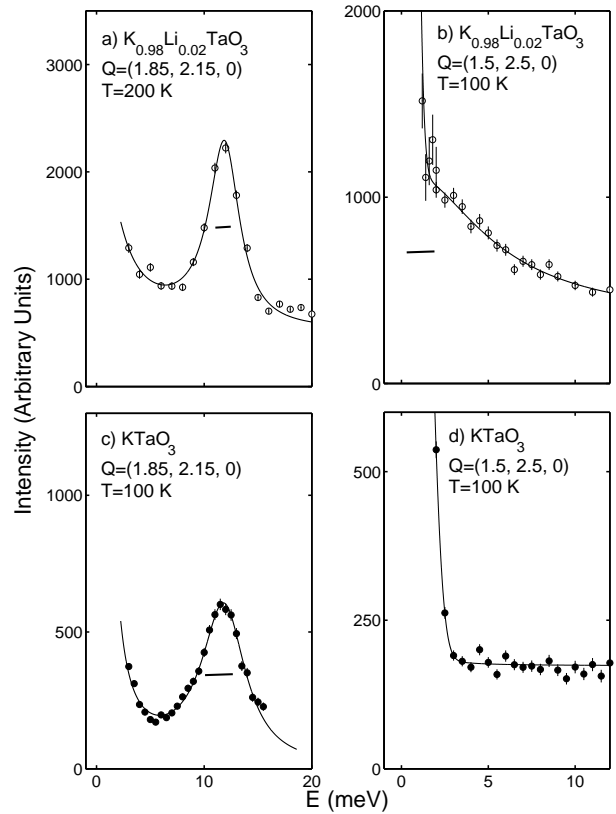}
\caption{\label{compare_phonon} Neutron inelastic scattering spectra from KLT(0.02) and KTaO$_3$ measured on the 1T thermal neutron spectrometer located at the LLB.  An underdamped TA mode is shown in panels $a)$ and $c)$ for both samples.  Energy broadened relaxational dynamics are observed in KLT (panel $b)$) where the intensity decreases monotonically with increasing energy transfer.  By contrast, KTaO$_3$ displays no such dynamics (panel $d)$) at the same wave vector and temperature.  The horizontal bars represent the instrumental elastic energy resolution FWHM (full-width at half-maximum).}
\end{figure}

KTaO$_3$ is a stoichiometric, parent compound of KLT for which the ferroelectric-active, low-energy TO mode softens monotonically with decreasing temperature,~\cite{Shirane67:57,Perry89:39,Axe70:1,Comes72:5,Farhi00:15} at least for all temperatures above $\sim 1$\,K as discussed in Ref.~\onlinecite{Wemple65:137}.  No long-range ordered structural phase transition, characterized by new, sharp (resolution-limited) Bragg peaks, has ever been observed.  For these reasons KTaO$_3$ is classified as an incipient ferroelectric.  It is known that replacing K$^+$ with Li$^+$ on the perovskite $A$-site stabilizes a ferroelectric ground state in KLT for sufficiently large $x > x_c$.  Microscopically, the much smaller Li$^+$ ion shifts away from the high symmetry position along one of six [100] directions thereby creating a local electric dipole, which has been studied using NMR techniques.~\cite{Klink83:27}  It has also been shown that Li$^+$ doping actually hardens the soft zone-center TO mode, which suggests that the transition to the ferroelectric phase is order-disorder in character and not displacive.~\cite{Vogt96:184,Vogt97:202}  However, consensus on the precise value of $x_c$ is lacking.  In 1981 Prater \emph{et al.} measured the Raman scattering from various KLT samples as a function of temperature and concluded that they were consistent with the presence of a tetragonal or orthorhombic ferroelectric phase for $x > 0.01$.~\cite{Prater81:23}  In 1987 Kleemann \emph{et al.} performed optical birefringence, refractive index, and dielectric permittivity measurements on KLT and proposed a crossover from a dipole-glass to ferroelectric phase at $x_c = 0.022$.~\cite{Kleeman87:04}  At the same time dielectric measurements have shown that the peak in the dielectric permittivity of KLT is broad in temperature and strongly dependent on frequency across a range of concentrations $0.01 \le x \le 0.035$ and thus similar to the dielectric response of PMN.~\cite{Samara03:15,Toulouse94:49}  By analogy with the lead-based systems, the elastic diffuse scattering in more heavily Li$^+$-doped KLT samples has been interpreted as resulting from short-range ferroelectric correlations or polar nanoregions.~\cite{Yong00:62}  These results are supported by earlier neutron studies of the TA phonons in KLT(0.035), which employed an external electric field to observe the splitting of the TA modes resulting from a cubic-to-tetragonal structural phase transition.~\cite{Toulouse94:49}  More recently a neutron scattering study by Wakimoto \emph{et al.} examined the lattice dynamics of a single crystal of KLT with $x=0.05$ and, just as in KTaO$_3$, found a zone-center TO mode that softens continuously on cooling and no evidence of any long-range ordered structural distortion.~\cite{Waki06:74}  Wakimoto \emph{et al.} also observed elastic diffuse scattering that increased monotonically on cooling below 200\,K, but it was present only near some nuclear Bragg peaks such as (210) and absent near others such as (100) and (200), and the diffuse scattering contours were elongated along $\langle100\rangle$; this situation differs markedly from that observed in PMN and PZN.  It is quite likely that some of the discrepancies between the aforementioned studies may be the result of uncertainties and/or gradients in the nominal values of the Li$^+$ concentrations.

While there have been extensive studies of KLT for $x > x_c$, there has been comparatively little effort for very low Li$^+$ concentrations for which a well-defined ferroelectric transition is absent.  For $x < x_c$ the dielectric constant displays a frequency-dependent and broad peak consistent with a relaxor phase, but no evidence of any elastic diffuse scattering nor any anomaly of the soft TO mode characteristic of this phase has been observed.~\cite{Yong00:62,Wen08:78}  Neutron experiments on KLT(0.02) have reported a strong enhancement of the nuclear Bragg peak intensity below a critical temperature that roughly matches the temperature at which the peak in the dielectric response occurs.  The Bragg peak enhancement is possibly indicative of a release of extinction and is similar to the behavior observed in PZN.~\cite{Stock04:69}  However no Bragg peaks were observed to split or distort, suggesting that the unit cell shape remains cubic at all temperatures.

The dielectric permittivity and crystal structure of KLT($x < x_c$) are very similar to those of PMN and PZN.  It is therefore puzzling that no other scattering signature of the relaxor phase, like those observed in lead-based systems, are seen in KLT($x < x_c$).  This begs the question of whether or not the dynamic features observed in the lead-based systems are common to other relaxor systems not based on PbTiO$_3$.  In this paper, we report neutron inelastic scattering results that demonstrate the existence of a definitive dynamic signature of localized (i.e. short-range ordered) ferroelectric correlations in KLT(0.02).  We observe strong, overdamped fluctuations that achieve maximum intensity at a temperature close to that at which the dielectric constant peaks and also where the nuclear Bragg peaks display a large enhancement in intensity.  These fluctuations occur on the THz timescale, which is much faster than those observed in the lead-based PMN and PZN, and are poorly correlated spatially.  A summary of these results is presented in Fig. \ref{compare_phonon}, which shows underdamped TA phonons for both KLT(0.02) and KTaO$_3$ (panels $a)$ and $c$)) and a broad relaxational mode in KLT (panel $b$)) that is absent in KTaO$_3$ (panel $d$)).

\section{Experimental details}

We use neutron scattering methods to study the structural dynamics of KLT and to connect the physics to the dielectric response.  Neutrons are uniquely well-suited to probe condensed matter systems over a broad range of momentum (length) and energy (time) scales with good resolution in both (typically $\Delta q \sim 0.01$\,\AA$^{-1}$ and $\Delta E \sim 1$\,meV or $\sim 0.25$\,THz).  This capability nicely complements dielectric measurements, which can only probe fluctuations near $|Q|=0$ but are sensitive to dynamics on much slower time scales ranging from GHz to mHz.  Therefore in order to understand how the fluctuations observed with dielectric measurements are spatially correlated and how they influence harmonic phonons on the THz timescale, it is essential to use neutrons, which are able to probe fluctuations throughout the Brillouin zone.

The KLT(0.02) single crystal used for this study is the same as that investigated in Ref.~\onlinecite{Wen08:78}.  It has dimensions $0.5 \times 1 \times 2$\,cm$^3$ and a room temperature lattice constant of 3.992\,\AA.  The Li concentration was determined by estimating the concentration in the melt and then verified from the position of the peak in the dielectric permittivity with temperature.  To test whether or not the scattering observed in KLT originated from Li$^+$ doping, we performed an independent series of measurements on a pure KTaO$_3$ crystal.  This crystal is the same as that used in Ref.~\onlinecite{Axe70:1}.

Most of the neutron scattering measurements reported here were performed on the 1T thermal-neutron triple-axis spectrometer located at the LLB-Saclay reactor.  The incident and final (scattered) neutron energies were defined through the use of a doubly-focused PG(002) monochromator and analyzer, which were used with open collimation.  For measurements on KLT(0.02) the final neutron energy was fixed to $E_f=14.7$\,meV and a highly-oriented pyrolytic graphite (HOPG) filter was used to remove higher-order neutrons from the scattered beam.  For elastic Bragg peak measurements an HOPG filter was also placed in the incident beam to further reduce the possibility of higher-order neutron contamination from the monochromator.  Our measurements were conducted primarily at low energies ($\le 12$\,meV) where the strong, overdamped signal is present.  To obtain a reciprocal space map of the lattice dynamics, we performed measurements on the NEAT cold-neutron chopper instrument, located at the HZB research reactor, using a fixed initial neutron energy $E_i=14.1$\,meV and the detector coverage on the neutron energy gain side.  A similar technique was used to map out the soft columns of scattering located near the zone boundaries in PMN.~\cite{Swainson09:79}

Measurements on pure KTaO$_3$ were made using the 1T spectrometer, where $E_i$ was fixed to 36\,meV using a Cu(111) monochromator.  A fixed and large incident energy was chosen to reduce background resulting from higher-order neutrons incident on the same area and hence improve sensitivity to any diffuse ``quasistatic" scattering near $E=0$.  The final neutron energy was scanned using the (004) reflection from a PG crystal analyzer, and the intensity was corrected for the $k_{f}^{3}/\tan(\theta_{analyzer})$ factor required to account for the change in resolution.~\cite{Shirane:book}  The sample was cooled using a closed-cycle cryostat.  Additional measurements of the TA phonon dispersion were performed using the BT9 thermal-neutron triple-axis spectrometer, located at the NIST Center for Neutron Research, using a PG(002) vertically-focused monochromator and a PG(002) analyzer.  The horizontal beam collimation sequence used on BT9 was $40$-80$'$-$S$-80$'$-$open$ ($S$ denotes the sample position).  An HOPG filter was inserted in the scattered beam to remove higher-order neutron contamination from the monochromator.  The final neutron energy was fixed to $E_f=14.7$\,meV and the energy transfer $E=E_i-E_f$ was scanned by varying the incident neutron energy.  We note that all data for both 1T and BT9 were corrected for contamination of the incident beam monitor as described in detail in the appendix of Ref.~\onlinecite{Stock2_04:69} and in Ref.~\onlinecite{Shirane:book}.  To verify consistency between the different experimental configurations we cross-checked the low-energy TA phonons shown in Fig.~\ref{dynamic_summary} $a)$ and $b)$.

\section{Elastic response and structural transition}

\begin{figure}[t]
\includegraphics[width=9cm] {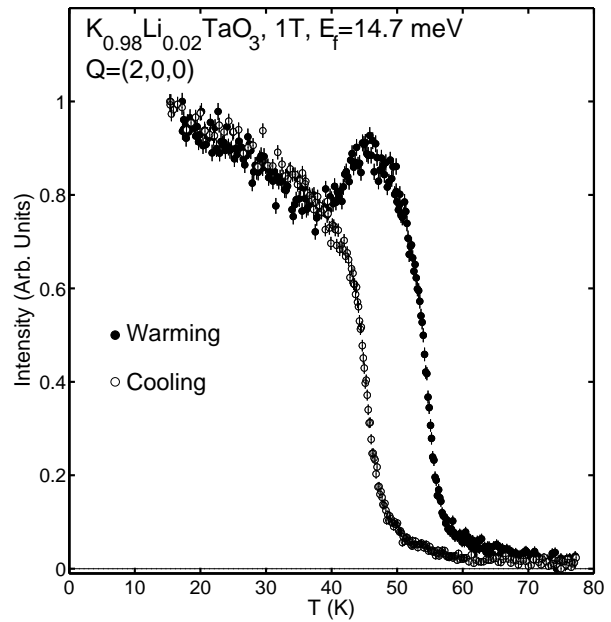}
\caption{\label{Bragg_T}  The change in Bragg peak intensity at $\vec{Q}$=(2,0,0) as a function of temperature measured on 1T (LLB, Saclay).  The open and filled circles correspond to cooling and warming sequences, respectively.  A large hysteresis is observed and the intensity displays a peak on warming.  The warming and cooling rates were set to 2\,K/min.}
\end{figure}

The structural properties of KLT(0.02) have been reported in detail by Wen {\em et al}.\ in both zero and non-zero applied electric field,~\cite{Wen08:78} but for completeness we review the zero-field properties relevant to the inelastic studies discussed below.  In the experiments described in Ref.~\onlinecite{Wen08:78} and here, no splitting of the Bragg peaks or any temperature-dependent anomaly in the lattice constants is seen.  This is consistent with the fact that the Li$^+$ concentration $x$ of our sample is less than the critical concentration $x_c$ (discussed earlier) required for a well-defined structural phase transition to occur.

We observe a very large increase in the Bragg peak intensity on cooling.  This is illustrated in Fig.~\ref{Bragg_T}, which shows the Bragg peak intensity at $\vec{Q}$=(2,0,0) as a function of temperature.  A large hysteresis of order 10\,K is seen, and on warming the intensity displays a distinctive peak near 50\,K, which is close to the temperature where the broad and frequency-dependent peak occurs in the dielectric permittivity.  The complex temperature dependence of the elastic scattering shown in Fig.~\ref{Bragg_T} is not expected for either a first or second-order phase transition and is similar to the loss of extinction in the Bragg peaks reported in some lead-based relaxors like PZN.~\cite{Stock04:69}   As noted in previous studies on KLT with small Li$^+$ concentrations (Refs.~\onlinecite{Wen08:78}), we do not observe any temperature-dependent elastic diffuse scattering that decorates the nuclear Bragg peaks as measured in Ref. \onlinecite{Yong00:62}.  The high-energy x-ray study of Wen \emph{et al.} did find evidence of diffuse scattering along $\langle100\rangle$ directions, but these were shown to be a result of the large energy resolution associated with the x-ray measurement, which integrated over low-energy phonons.  Our measurements here have all been performed with a monochromator and analyzer, which provide an elastic ($E=0$) energy resolution of $\sim 1$\,meV full-width at half maximum (FWHM).  Therefore, we can state with confidence that KLT(0.02) does not undergo a bulk, long-range ordered, structural phase transition, but it does exhibit an anomaly that results in a loss of extinction (higher intensity) in the Bragg peaks on cooling to low temperature.  The origin and the nature of this is discussed later in the text and related to the dielectric response.

\section{Low-energy phonon dispersion - response of the TA$_1$ phonon to Li$^+$ doping}

\begin{figure}[t]
\includegraphics[width=8.9cm] {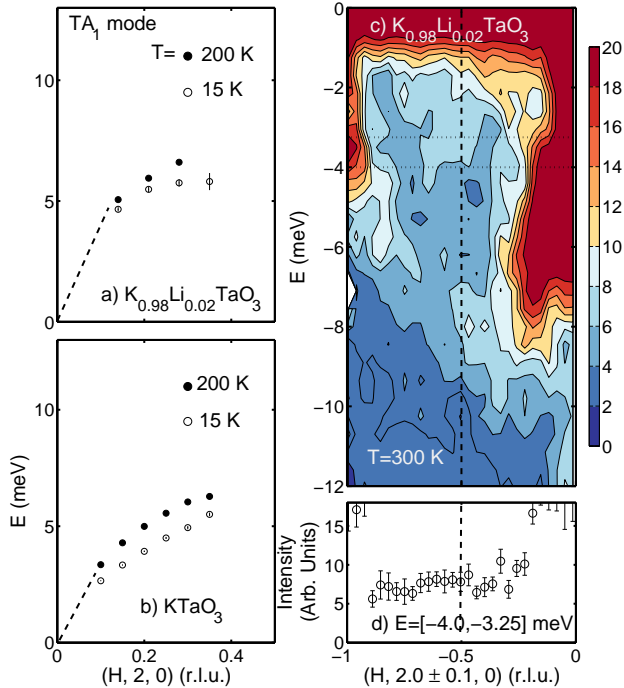}
\caption{\label{dynamic_summary}  Panels $a)$ and $b)$ show the low-energy TA$_1$ phonon peak positions for KLT(0.02) and pure KTaO$_3$, respectively.  A constant-$\vec{Q}$ map of the low-energy phonons measured on NEAT is illustrated in panel $c)$ at 300\,K that spans two Brillouin zones (divided by the dashed line).  The negative values of the energy transfer $E$ indicate that the experiment was performed on the neutron energy gain side.  The region of high-intensity (shown in red) at $E=0$ results from elastic incoherent scattering; the high intensity regions near $\vec{Q}=(-1,2,0)$ and (0,2,0) are due to the low-energy TA and TO phonons.  Of particular interest for this study is the weaker region of intensity that exists between the well-defined acoustic modes at the two zone centers, which displays a weakly varying intensity ridge. $d)$ shows a constant energy cut illustrating that there is no strong momentum dependence to this weak scattering throughout the Brillouin Zone.}
\end{figure}

In this section we review the low-frequency lattice dynamical response of KLT(0.02) and KTaO$_3$ as a function of temperature.  The low-energy TA and TO modes have been the topic of several studies for KLT(0.02) (Ref.~\onlinecite{Wen08:78}) and KTaO$_3$ (Ref.~\onlinecite{Axe70:1,Comes72:5,Farhi00:15}).  We note that a detailed neutron scattering study of the TA phonons in a more highly doped sample of KLT(0.035) found clear evidence of a structural distortion through the use of an electric field.~\cite{Toulouse94:49}  We performed surveys of the low-energy TA and TO phonons, and our results are in agreement with previous descriptions.


Fig.~\ref{dynamic_summary} $a)$ and $b)$ shows the dispersions of the TA$_1$ modes (transverse acoustic phonons that are polarized along [100] and propagating along [010]) for  KLT(0.02) and KTaO$_3$, respectively, at 15\,K and 200\,K.  In agreement with the classic mode-coupling analysis of Axe \emph{et al.},~\cite{Axe70:1} we find that the TA$_1$ mode energy decreases on cooling for both compounds, which reflects the softening of the TO mode.  And while the energy scales of the TA$_1$ phonon branches in KLT(0.02) and KTaO$_3$ are similar, there are two important features to note here.  First, the long-wavelength modes in KTaO$_3$ are noticeably lower in energy.  This is consistent with the mode-coupling scenario and the fact that Li-doping stiffens the TO branch in KTaO$_3$.~\cite{Vogt96:184,Vogt97:202}  Second, the relative change in the TA$_1$ mode energy with temperature in KLT(0.02) is slightly less than in KTaO$_3$.  This suggests that the TO-TA mode coupling is weaker in KLT(0.02), which is again consistent with the Li-induced stiffening of the TO branch.  This situation bears an interesting similarity to the case of PMN versus PbTiO$_3$, which was studied by Stock \emph{et al.} who demonstrated that the TO-TA mode coupling in PMN is extremely weak and thus cannot be the cause of the waterfall effect.~\cite{Stock05:74,Hlinka03:91}  These phonon data therefore confirm the elastic results of Fig. \ref{Bragg_T} in that both show that the Li$^+$ dopants have a significant effect on the bulk properties of the KLT(0.02) crystal.

\begin{figure}[t]
\includegraphics[width=9cm] {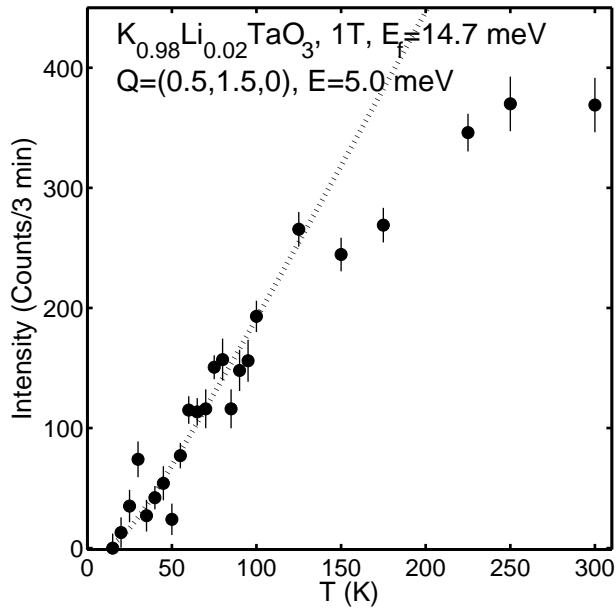}
\caption{\label{zone_boundary_T}  Temperature dependence of the change in scattered intensity at the zone-boundary $M$-point $\vec{Q}=(0.5,1.5,0)$ at an energy transfer of 5\,meV.  The data were measured by tuning the spectrometer to the $M$ point and taking the lowest temperature as the background.  The dashed line represents the Bose factor at each temperature scaled to fit the temperature dependence below $\sim 125$\,K.}
\end{figure}

To search for other effects of the Li$^+$ dopant on the lattice dynamics that might be relevant to the elastic anomalies discussed in the previous section, we performed a reciprocal-space survey of the lattice dynamics of KLT(0.02) using time-of-flight neutron scattering methods.  Fig.~\ref{dynamic_summary} $c)$ illustrates an intensity map generated by a series of constant-$\vec{Q}$ slices that were measured using the NEAT direct-geometry time-of-flight spectrometer (HZB, Berlin) on KLT(0.02).  The experiment was performed using the neutron energy gain-side (energy transfer $E=E_i-E_f \le 0$) at 300\,K because the energy resolution is narrower near the elastic ($E=0$) position and progressively worsens at higher energy transfers; this is a configuration that mimics that of a triple-axis spectrometer and better matches the characteristics of the cross section being measured.  The region of high intensity around the elastic position results from the relatively large incoherent cross section of potassium.  The lines of high intensity at $\vec{Q}$=(0,2,0) and (1,2,0) represent strong scattering from the low-energy TA and TO modes previously characterized.  An especially interesting feature is the significant inelastic scattering seen near the zone-boundary X-point $\vec{Q}$=(-0.5,2,0), which extends from $E=0$ to at least $\sim -4$\,meV, close to the top of the acoustic TA$_1$ phonon branch.  While the scattering is most obvious at the zone boundary, where the acoustic modes lie at a higher energy (Fig. \ref{dynamic_summary} $a)$ and $b)$), the scattering is not well defined in momentum as evidenced by the extent of the contours with respect to the size of the Brillouin zone. This inelastic scattering is broadly distributed in $H$ and, in this respect, does not strongly resemble the so-called ``columns" of scattering reported by Swainson \emph{et al.} in PMN using similar techniques.~\cite{Swainson09:79}  The fact that the scattering is highly extended in momentum implies the presence of a highly localized object in real space.  To explore this broad scattering in more detail, we performed additional measurements using a thermal-neutron triple-axis spectrometer which we discuss in the following section.

\section{Overdamped low-energy fluctuations}

Motivated by the unusual inelastic scattering shown in Fig. \ref{dynamic_summary} and the prior discovery of columns of spectral weight (i.\ e.\ scattering that is broadly distributed in energy, but comparatively narrower in momentum) in PMN located at the $M$- and $R$-point zone boundaries,~\cite{Swainson09:79} we performed additional measurements of the dynamics in KLT(0.02) using thermal-neutron triple-axis techniques.  Fig.~\ref{zone_boundary_T} shows how the inelastic scattering intensity measured at the zone boundary $M$-point $\vec{Q}=(0.5,1.5,0)$ at an energy transfer of 5\,meV changes as a function of temperature.   The low-temperature value is assumed to be the background.  The dashed line represents the Bose factor, which gives the temperature dependence expected for a simple thermal excitation.  The measured temperature dependence is clearly inconsistent with that of the Bose factor.  Thus we investigated this scattering further in momentum, energy, and temperature.

\begin{figure}[t]
\includegraphics[width=8.5cm] {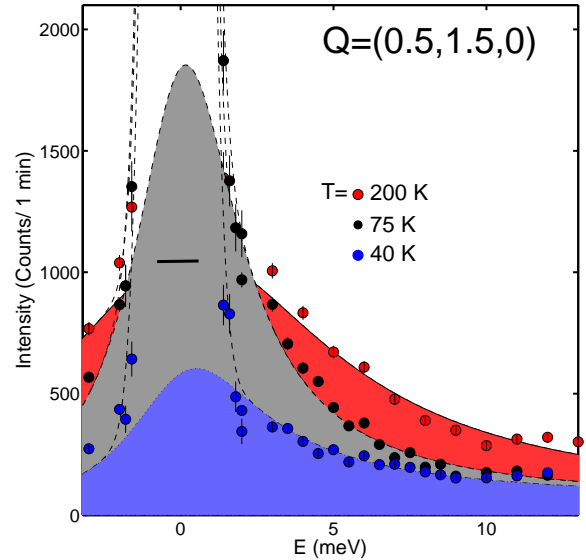}
\caption{\label{constantQ_T}  Constant-$\vec{Q}=(0.5,1.5,0)$ scans measured with the 1T thermal-neutron spectrometer with $E_f=14.7$\,meV.  Panels $a)$-$c)$ shows scans at 40, 75, and 200\,K respectively.  The solid line is a fit to the relaxational lineshape described in the text.  The dashed line is a temperature-independent Gaussian function used to describe the strong incoherent elastic scattering.  The horizontal bar represents the experimental elastic energy resolution (full-width at half maximum).}
\end{figure}

Fig.~\ref{constantQ_T} displays a series of constant-$\vec{Q}$ scans measured at $\vec{Q}=(0.5,1.5,0)$ at 40, 75, and 200\,K.  With increasing temperature a broad ``quasielastic" component, indicated here by the solid line centered at $E=0$, is observed in addition to the strong, narrow in energy, incoherent elastic component, which dominates the total scattering near $E=0$.  This quasielastic scattering contrasts sharply with that from a well-defined harmonic mode such as a phonon, for which the lineshape in energy would be characterized by an underdamped peak centered at a non-zero energy.  The observed excitations are therefore overdamped and display an energy linewidth (which is inversely related to the excitation lifetime) that is considerably broader than the intrinsic resolution of the spectrometer.

To parameterize these broad and overdamped fluctuations, we have fit the constant-$\vec{Q}$ scans to the following modified Lorentzian function, which is characteristic of a relaxational lineshape with a single energy scale:

\begin{eqnarray}
I(E)=\chi_{0}[n(E)+1] \times {E \over {1+\left({E\over \gamma}\right)^{2}}}.
\label{mod_lor}
\end{eqnarray}

\noindent For completeness we have also performed the analysis using a damped harmonic oscillator, and this is presented in the appendix.  We find consistent results between the two different lineshapes, and given the fact that the modified Lorentzian has fewer free parameters we use this lineshape for the remainder of the paper.  Fitting was done using two parameters plus a temperature-independent Gaussian function centered at the elastic $E=0$ position to account for the incoherent elastic scattering from the sample and mount.  The parameter $\chi_{0}$ is an overall amplitude and can be related to the real part of the susceptibility via the Kramers-Kronig relation.  The parameter $\gamma$ describes the linewidth of the modified Lorentzian function and is related to the relaxational time via $\gamma \propto 1/\tau$, and $[n(E)+1]$ is the Bose factor.  The same lineshape has been used to describe the quasielastic scattering observed in the relaxor PMN (Ref.~\onlinecite{Gvas04:69}) as well as in various spin-glass systems (Ref.~\onlinecite{Maletta81:52}).

The quality of the fits to the data using Eqn.~\ref{mod_lor} are illustrated by a series of representative curves in Fig.~\ref{constantQ_T} as well as in Fig.~\ref{constantQ_bz_near} and Fig.~\ref{constantQ_bz}, which are discussed below.  The solid line is a fit to Eqn.~\ref{mod_lor} and the dashed line represents the effective background given by the Gaussian function used to describe the elastic incoherent scattering.  We see that the model cross section does an excellent job describing the data over a broad temperature range.

\begin{figure}[t]
\includegraphics[width=9cm] {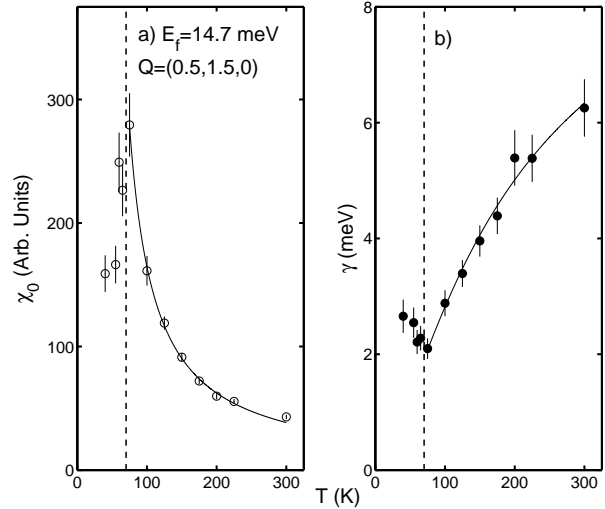}
\caption{\label{parameters}  The parameter $\chi_{0}$ and relaxation rate $\gamma$ are plotted in panels $a)$ and $b)$ respectively.  The parameters were extracted from constant-$\vec{Q}$ scans measured using the 1T thermal triple-axis spectrometer at $\vec{Q}=(0,1.5,0.5)$.  The vertical dashed line indicates the peak position of the frequency-dependent dielectric response at $\sim$ 200 kHz. The solid lines are fits to a Curie form $(a)$ and a Arrhenius law $(b)$.}
\end{figure}

This analysis employs only two temperature-dependent parameters, which are represented in Eqn.~\ref{mod_lor} by $\chi_{0}$ and $\gamma$.  The temperature dependence of both parameters is plotted in Fig.~\ref{parameters}.  The vertical dashed line in Fig.~\ref{parameters} is the average peak position of the real part of the dielectric permittivity ($\epsilon'$) over the frequency range 100\,Hz-1\,MHz shown in Fig.~1 in Ref.~\onlinecite{Wen08:78} (which examined the same KLT(0.02) sample studied here).   A clear peak is observed in $\chi_{0}$ at this temperature that is accompanied by a minimum of the energy linewidth $\gamma$ at $\sim 2$\,meV.  The peak in the susceptibility is also near where a large increase is observed in the Bragg peak intensity with decreasing temperature (Fig. \ref{Bragg_T}).   We note that $\gamma$ measures the half-width of the scattering; therefore the quasielastic scattering always remains considerably broader than the energy resolution of the spectrometer, which is $\Gamma=0.55$\,meV (half-width at half-maximum).  The response is not critical in nature as is characteristic of a long-range ordered structural phase transition.  If this were the case, then a divergence of $\chi_{0}$ and $1/\gamma$ would be manifest (see, for example, Fig.~10 in Ref.~\onlinecite{Stock11:83}, where a similar analysis was performed for the magnetic transition in Ba$_3$NbFe$_3$Si$_{2}$O$_{14}$).  The analysis shows that the broad quasielastic scattering shown in Fig.~\ref{constantQ_T} is associated with dielectric properties and the broad peak in the dielectric response.  To better confirm this point, we have performed measurements on KTaO$_3$ and find this scattering to be absent in the undoped, parent material.  The anomaly of the energy linewidth near 70 K in Fig.~\ref{parameters} $b)$ is reminiscent of the dynamics in the magnetic dynamics in the disordered multiferroic PbFe$_{1/2}$Nb$_{1/2}$O$_3$.~\cite{Stock13:88}

While our measurements at non-zero momentum transfer are, in principle, different from the dielectric response associated with $|Q|$=0, the connection between the peak in the dielectric constant and the anomalies observed $\sim$ 60-70 K in $\gamma$ and $\chi_{0}$ discussed above suggest a common origin.  We note, however, that the quasistatic scattering we observe is present across the Brillouin zone include the zone center and boundary.  We cannot rule out the presence of zone boundary distortions and in the lead based relaxors, zone boundary and center phonon anomalies are observed at the same temperature.~\cite{Swainson09:79}

To compare our results directly we have fit (represented by the solid lines in Fig. \ref{parameters}) the high temperature dependence of $\chi_{0}$ to a Curie form $\chi_{0} \propto 1/(T-\Theta_{W})$ and $\tau=\tau_{0}\exp\left({U/k_{B}(T-\Theta_{W})}\right)$ (noting $\gamma \propto 1/\tau$ and $U$ is an activation energy). The origins of such a ``law" are described in Ref. \onlinecite{Saslow88:37}.  We have included a ``Weiss" temperature in both fits to $\chi_{0}$ and $\gamma$ to improve the fit and to account for the anomaly at $\sim$ 60-70 K.   We note that our data and temperature range are not able to distinguish between a power law ($\tau \propto 1/(T-\Theta_{W})$ suggested for spin glasses ~\cite{Aeppli85:54,Yamada74:9}) and an Arrhenius form used here.  We have chosen the exponential Arrhenius form to compare the energy scale of the fluctuations with dielectric measurements.   Based on this analysis we derive a Weiss temperature of $\Theta_{W}$=38 $\pm$ 3 K, $U$=187 $\pm$ 20 K, $\tau_{0}$=0.10 $\pm$ 0.02 ps.   These values are very different from those derived from dielectric measurements where the energy barrier for the relaxational ``site-hopping" process are $U\sim$ 1000-2000 K and $\tau_{0}\sim$ 0.01 ps and associated with dynamics of the Li$^+$ ions.~\cite{Pattnaik00:62}  The Curie temperature $\Theta_{CW}$ is similar to the glass temperature measured with NMR and close to where a quadrupole split spectrum appears.~\cite{Klink84:30,Klink83:27}  However, given such disparate energy scales derived from Arrhenius fits, we conclude that the observed dynamics here are not directly associated with Li$^+$ relaxational processes.   To address this point and the origin of the dynamics, we now discuss the momentum dependence.

\begin{figure}[t]
\includegraphics[width=8.5cm] {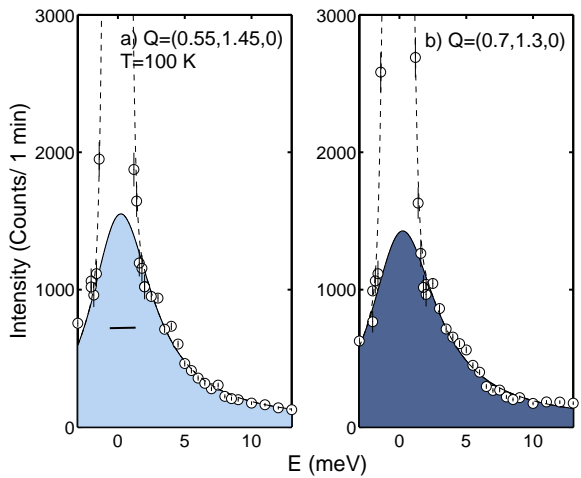}
\caption{\label{constantQ_bz_near}  Constant-$\vec{Q}$ scans performed at two wave vectors near the $M$-point Brillouin zone boundary located at $\vec{Q}$=(0.5, 1.5, 0). The values of the linewidth, $\gamma$, in panels $a)$ and $b)$ are are $3.0 \pm 0.2$\,meV and $2.4 \pm 0.2$\,meV, respectively.}
\end{figure}

The quasielastic scattering represented by a relaxational lineshape is not only broad temporally (in energy), but it is also spatially uncorrelated and hence displays very weak momentum dependence.  This is illustrated in Fig.~\ref{dynamic_summary} by the broad scattering observed throughout the Brillouin zone.  We explore this point further in Fig.~\ref{constantQ_bz_near}, which shows a series of constant-$\vec{Q}$ scans near the $\vec{Q}$=(0.5,1.5,0) zone boundary studied in detail above in Figs.~\ref{constantQ_T} and \ref{parameters}.  Scans for $\vec{Q}=(1-q,1+q,0)$ over the range of $q=[0.5,0.25]$ are shown at $T=100$\,K.  The quasielastic scattering is clearly observed over this broad range in momentum transfer across a single Brillouin zone.  The linewidth $\gamma$ also appears to decrease away from the zone boundary and at small values of $|Q|$ (the values of $\gamma$ are given in the figure caption).  The overdamped nature of this quasistatic scattering and the broadening with momentum transfer does not represent a dispersion (as observed for the acoustic harmonic phonons as in Fig. \ref{dynamic_summary}).

\begin{figure}[t]
\includegraphics[width=9.25cm] {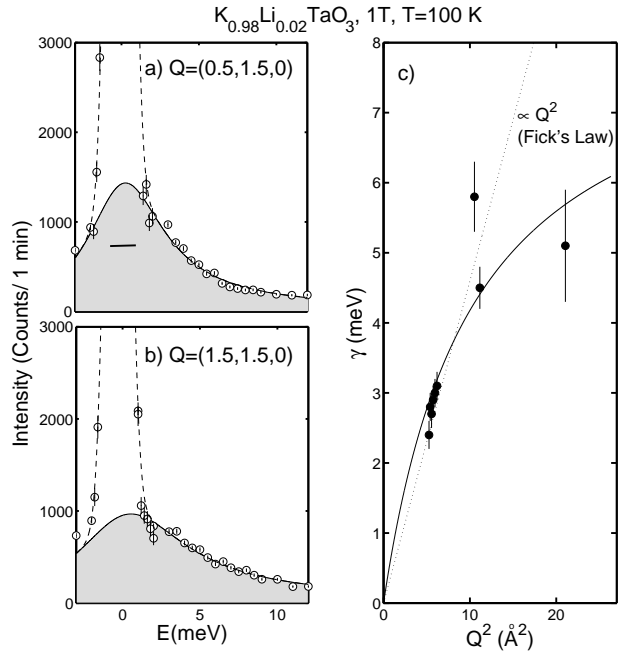}
\caption{\label{constantQ_bz} $a)$ and $b)$ show constant-$\vec{Q}$ scans measured at 100\,K for $\vec{Q}=(0.5,1.5,0)$ and (1.5,1.5,0).  The fitted linewidths are $\gamma=3.1 \pm 0.2$\,meV and $4.5 \pm 0.3$\,meV respectively.  $c)$ shows a plot of the half-width ($\gamma$) as a function of $Q^{2}$.  The dashed line is a fit to Fick's law for diffusion, and the solid line is a fit to a heuristic model for oscillating motion.}
\end{figure}

We next measured the scattering at different $M$-points, shown in Fig.~\ref{constantQ_bz} $a)$ and $b)$, as well as at different $|Q|$, shown in Fig.~\ref{constantQ_bz} $c)$.  Fig.~\ref{dynamic_summary} shows the scattering near the $X$-point $\vec{Q}=(0.5,2,0)$, whereas the temperature dependent measurements were made around the $M$-point $\vec{Q}=(0.5,1.5,0)$, shown in Fig.~\ref{constantQ_T}.  Fig.~\ref{constantQ_bz} shows analogous scans of the relaxational quasielastic scattering at $T=100$\,K.  Panel $a)$ and $b)$ show scans performed at the $M$-points $\vec{Q}=(0.5,1.5,0)$ and $(1.5,1.5,0)$, respectively.  The different values of the fitted energy linewidth $\gamma$ are given with error bars in the figure caption.  The quasielastic scattering is always present and broadens in energy with increasing $|Q|$.  Measurements at equivalent Brillouin zone positions and at different $|Q|$ confirmed this trend.  This proves that the scattering does not follow the symmetry constraints of the parent lattice and  indicates the origin as highly localized objects.  A discussion of the relation between localized polar regions and the diffuse scattering and dielectric measurements is presented in Ref. \onlinecite{Vug06:73}.

The above analysis yields three key results.  First, it demonstrates the presence of broad, quasielastic scattering that peaks (measured through $\chi_{0}$ and $\gamma$ in Fig.~\ref{parameters}) at roughly the same temperature as does the dielectric permittivity and also where a sharp increase in the Bragg peak scattering intensity is observed.  The quasielastic scattering is overdamped, and cannot be characterized as a well-defined dispersing harmonic mode.  Second, the linewidth of the scattering increases with $|Q|$, but it does not follow the symmetry operations expected given the Brillouin zone fixed by the parent KTaO$_3$ lattice.  Third, the activation energy ($U$) and the characteristic timescale ($\tau_{0}$) are both an order of magnitude smaller than that measured in the dielectric response.  Based on these three results, we suggest that the scattering is associated with dilute, highly local droplets of dipoles, or ferroelectric regions, created by the introduction of Li$^+$ dopants, which enter randomly into the lattice.  Given the ferroelectric nature of the localized object, we would expect a response at the zone center, consistent with dielectric measurements.  Unfortunately the non-zero energy and momentum resolution of the neutron spectrometer make measurements at the zone center problematic due to contamination from nearby acoustic modes.  However, by extrapolating the results obtained at non-zero wave vectors, we are able to infer that a response exists at the zone center that is consistent with the ferroelectric nature postulated here.  The dynamics cannot be directly related to the hopping of Li$^+$ cations given the large discrepancy between the activation energies derived here and from dielectric measurements.  This is further confirmed by the fact that the integrated intensity of the relaxational ``quasi static" component (Fig.~\ref{compare_phonon}) is comparable to that in the low-energy acoustic phonons, which makes it unlikely to originate from the local hopping of a 2\% dopant.  These local dipoles are highly dilute and spatially uncorrelated as is evident from the absence of any well-defined, long-range-ordered, ferroelectric structural transition.  The comparison of the dielectric and neutron data performed on different timescales indicates a hierarchy of degrees of freedom in KLT(0.02) with different energy scales.  Such a scenario has been discussed in relation to the dynamics of glasses.~\cite{Palmer53:84}

The scattering is unlikely to originate from a strictly zone-boundary distortion (such as tilting of the octahedra) because the temperature dependence of the neutron scattering linewidth and susceptibility display anomalies at the same temperature as do the $|Q|=0$ dielectric measurements (Ref.~\onlinecite{Doussineau93:24}) and the Bragg peaks presented in Fig.~\ref{Bragg_T}.  However, as noted by Swainson {\em et al}.,~\cite{Swainson09:79} the lead-based relaxors display both zone center and zone boundary phenomena.  It should also be noted that the response observed here is proximate to zone-center diffuse scattering for slightly larger Li$^+$ concentrations (Ref.~\onlinecite{Waki06:74}).  Besides the connection to the dielectric data, this behavior is also substantiated by calculations that suggest that the distortion is a zone-center response (Ref.~\onlinecite{Tupitsyna01:64}) and also ultrasound measurements which probes the acoustic velocity in the limit as $q$ approaches the zone center (Ref. \onlinecite{Doussineau93:21}).

We now discuss the momentum dependence of the energy linewidth plotted in Fig.~\ref{constantQ_bz} $c)$, which shows the half-width $\gamma$ as a function of $|Q|^{2}$.  We emphasize that we observe the fluctuations to be overdamped (the linewidth being larger than the energy position) at all momentum transfers and temperatures studied.  The half-width $\gamma$ is well-described by a $Q^{2}$ law for small momentum transfers, but this behavior seems to breakdown at larger momentum transfers taken in higher Brillouin zones.  Following the study of localized dynamics in molecular systems (Ref.~\onlinecite{Teixeira85:31}), we fit our data to the following heuristic model, which has been applied to diffusing molecules for which the local oscillation time is much longer than the time for translational diffusion.

\begin{eqnarray}
\gamma=\gamma_{0}\left(1 - {1\over{1+\alpha Q^{2}}} \right).
\label{linewidth}
\end{eqnarray}

\noindent Here we assume that the Debye term $e^{-2W} \sim 1$ because the quality of the data in Fig.~\ref{constantQ_bz} does not support introducing a third parameter into the data analysis.  The parameter $\gamma_{0}$ is inversely proportional to the oscillation time and $\alpha=D\tau_{0}$, where $D$ is the diffusion constant.  In the limit of small momentum transfer Eqn.~\ref{linewidth} reduces to Fick's law (shown by the dashed line in Fig.~\ref{constantQ_bz} $c)$), which states that the linewidth broadens in proportion to $|Q|^2$.  While the large momentum transfer data are somewhat scattered in Fig.~\ref{constantQ_bz} $c)$, the data do show a broadening of the lineshape consistent with localized, and spatially uncorrelated dipoles.  The localized nature prevents them from having the same periodicity as the lattice and therefore prevents the dynamics from being concentrated near the zone center.  We speculate that  this ``diffusion" is tied to the fluctuations of locally ferroelectric distorted regions in a paraelectric background instead of the actual physical hopping of Li$^+$ cations.  We emphasize that, given the overdamped nature of the fluctuations, this is not a dispersion as is defined for harmonic fluctuations (as shown for the acoustic phonons in Fig.~\ref{dynamic_summary}).

\section{Conclusions and discussion}

We have presented a neutron inelastic scattering study of the low-energy lattice dynamics in KLT(0.02).  This compound undergoes no long-range ordered structural distortion; instead it displays a strongly frequency-dependent dielectric permittivity that peaks broadly in temperature, which is very similar to that observed in the lead-based relaxor ferroelectrics.  The main finding of our study is the existence of a broad quasielastic component that peaks in temperature near to where an anomaly is observed in the dielectric response.  This scattering is well described by a relaxational lineshape that is characterized by a single energy scale $\gamma \propto 1/ \tau$.

The relaxational dynamics we report here are tied to local ferroelectric correlations induced by the Li$^+$ dopants.  We base this conclusion on several experimental facts.  First, the scattering we observe in KLT(0.02) is absent in pure KTaO$_3$, which displays no dielectric anomaly at any temperature.  Second, the spectral weight associated with the relaxational dynamics in KLT(0.02), probed through $\chi_{0}$ in Eqn.~\ref{mod_lor}, peaks at the same temperature near to which the nuclear Bragg peak intensity suddenly increases and where a peak (albeit frequency-dependent and broad) is observed in the dielectric response.  Third, the quasielastic scattering does not follow the symmetry of the KTaO$_3$ lattice; this indicates that it originates from local objects created by the dilute Li$^+$ dopants.

The presence of a broad relaxational component to the scattering in KLT(0.02) is corroborated by Raman measurements, which probe dynamics near the zone center.  Prater {\em et al}.\ reported evidence of new excitations around $50$\,cm$^{-1} \sim 6$\,meV that broaden considerably with temperature in KLT(0.014).~\cite{Prater81:23}  Our results show that these excitations are broad in momentum and thus consistent with local (uncorrelated) objects.   For larger concentrations of Li$^+$ where a ferroelectric ground state is present, the local dipoles become spatially correlated and result in a soft TO mode that hardens at low temperatures and a subsequent splitting of the TA and TO phonons.~\cite{Toulouse94:49,Waki09:600,Waki06:74}

It is interesting to compare our results on KLT(0.02) with those obtained from the lead-based relaxor ferroelectrics represented by PbMg$_{1/3}$Nb$_{2/3}$O$_3$  (PMN).  PMN displays very strong and temperature-dependent diffuse scattering that is somewhat spatially correlated.  While the diffuse scattering in PMN is also broadly distributed in momentum, it is also highly structured and has been described in detail in a number of x-ray and neutron studies.  The scattering observed in KLT(0.02) is comparatively uncorrelated in momentum and better described by local, uncorrelated objects.  The dynamics of KLT(0.02) are also significantly faster than those associated with the diffuse scattering observed in PMN.  A neutron spin-echo study performed on PMN by Stock {\em et al}.~\cite{Stock10:81}) found excitations with a timescale of order $\sim$GHz.  In KLT(0.02) we observe dynamics on the order of 1\,meV or $\sim$THz.  The activation energies derived from the temperature dependence for PMN and KLT(0.02) are $U\sim 1100$\,K and $187 \pm 20$\,K, respectively.  There are therefore two key differences between the quasielastic diffuse scattering present in PMN and KLT(0.02):  (1) The scattering in KLT(0.02) is spatially uncorrelated in comparison to the strongly momentum-dependent and anisotropic diffuse scattering observed in PMN, and (2) the relaxational dynamics in KLT(0.02) are roughly two to three orders of magnitude faster than those in PMN.

The relaxational or ``quasistatic" component observed here in KLT(0.02) near the elastic ($E=0$) line contrasts strongly with the harmonic soft mode anomaly observed in KLT at larger Li$^+$ concentrations.  These two rather disparate responses can be reconciled through the interpretation of the relaxational dynamics as a ``central" component (i.\ e.\ one centered around $E=0$) like that first discovered in SrTiO$_3$ by Shapiro {\em et al}.~\cite{Shapiro72:6} and discussed theoretically by Halperin {\em et al}.~\cite{Halperin76:14} in terms of slowly-relaxing defects.  Similar ideas have also been applied to doped cuprate magnets,~\cite{Stock06_2:73} where, in the small doping limit, slowly-relaxing defects produce a central component that gives rise to relaxational dynamics like those described for KLT(0.02).  The lack of a strong momentum dependence indicates weak doping.  For larger doping concentrations the defects couple to the harmonic mode to produce a well-defined phase transition more characteristic of what is observed in the lead-based relaxors.  Although we draw this comparison, it is important to note that typical central peaks in the literature have energy widths that are comparable to the instrumental resolution and phonon linewidths.  The linewidths observed for KLT(0.02) are significantly broader, indicating a weak doping limit.  The relaxational lineshape has some strong similarities to the dynamics observed near the zone center in KNbO$_3$, however the fluctuations in KNbO$_3$ are slower, having energy widths less than $\sim 1$\,meV, and comparatively well-correlated in momentum.~\cite{Gvas04:80}

The structural fluctuations in KLT(0.02) soften on cooling, yet no long-range ordered phase transition is observed with diffraction and there is no evidence of a soft TO phonon anomaly like that in PMN or PbTiO$_3$.  Instead we observe only a strong increase in the Bragg peak intensity.  We suggest that the low-energy, local, polar domains that we have associated with these fluctuations freeze near $\sim 40$\,K, but that they do so in a random fashion given the lack of any strong momentum dependence.  This likely creates a random local field that could then explain the history dependence and slow frequency dependence observed near the transition temperature.  The Ising nature of the Li$^+$ displacement preserves the broad peak in the dielectric response and prevents the random field from entirely washing out the transition.~\cite{Imry75:35,Leheny03:32}  This scenario is consistent with studies of model Ising magnets in a random field, for which strong hysteresis and memory effects are also observed.~\cite{Hill91:66,Hill97:55}  This crossover from glassy (low-energy dynamics) to random-field behavior in KLT(0.02), which is characterized by a broadened dielectric peak and hysteresis, shares strong similarities with dilute ferromagnets including the lack of sharp propagating modes, the momentum dependence for $\gamma$, also the temperature dependence.~\cite{Aeppli84:29}  Given that random fields play a key role in relaxor ferroelectrics, and given the close analogy with magnetic systems, we refer to KLT(0.02) as an incipient relaxor with properties that are more consistent with those of fluctuating glasses than with random fields.

In summary, we have measured the dynamics associated with local dipoles created by the substitution of Li$^+$ for K$^+$ in the incipient ferroelectric perovskite KTaO$_3$.  The low Li$^+$ content of 2\% is insufficient to disturb the cubic ground state structure.  The dipoles are found to fluctuate on a timescale of $\sim$THz, which is several orders of magnitude faster than that observed in the lead-based relaxors.  The dipole dynamics, characterized using quasielastic neutron scattering methods, are broadly distributed in momentum and thus consistent with uncorrelated, spatially-localized objects.  Given (1) the absence of any static diffuse scattering, (2) the local nature of the dipoles, and (3) the significantly faster dynamics, we suggest that KLT at these low Li concentrations not be classified as a relaxor, but rather as an incipient relaxor by analogy with the lack of ferroelectricity in KTaO$_3$.


\section{Acknowledgments}
C.S. is grateful to the Carnegie Trust for the Universities of Scotland for financial support during this work.  Research at the Oak Ridge National Laboratory for one author (L.A.B.) is sponsored by the U.S. Department of Energy, Basic Energy Sciences, Materials Sciences and Engineering Division.

\section{Appendix - Damped harmonic oscillator description of the inelastic response}

The low-energy scattering described in the main text is heavily damped and characterized by a linewidth that is always considerably larger than the instrumental energy resolution.  This is evident from Fig.~6 where the full-width at half-maximum ($2\times \gamma$) of the scattering intensity is greater than 4\,meV.  By comparison the instrumental energy resolution at the elastic line is measured and calculated to be 1.5\,meV.  Although the inelastic response is broad in energy, it does have a tendency to be peaked at very low energies (see Figs.~7 and 8).  This situation is similar to that encountered in previous studies of amorphous magnets (Ref.~\onlinecite{Maletta81:52}) where two model cross sections were found to give equally good fits and physical interpretations of the data: 1) the ``modified Lorentzian" (used here) and 2) the damped harmonic oscillator.  Thus, while we favor the single modified Lorentzian description, we will discuss the alternate description here and compare the results of this analysis to those presented in the main text so that the reader will be able to draw his/her own conclusions.

\begin{figure}[t]
\includegraphics[width=9.25cm] {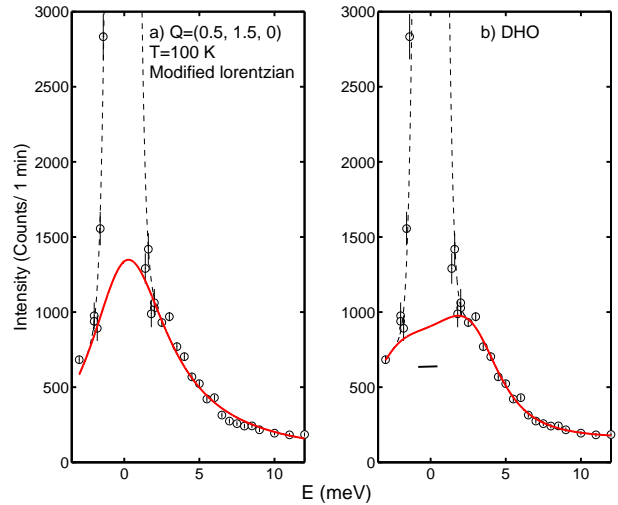}
\caption{\label{compare_lineshape} Constant-$\vec{Q}=(0.5,1.5,0)$ scan measured at 100\,K fit to a) a modified Lorentzian (Eqn.~\ref{mod_lor}) and b) a damped harmonic oscillator (DHO - Eqn. \ref{DSO}) lineshape.  The horizontal bar in panel b) represents the instrumental elastic energy resolution (FWHM).}
\end{figure}

The neutron inelastic scattering response $I(\vec{Q},E)$ is proportional to $S(\vec{Q},E)$, where $S(\vec{Q},E)$ is directly proportional to the imaginary part of the susceptibility ($\chi''$) times the Bose factor [$n(E)+1$],

\begin{eqnarray}
I(\vec{Q},E)\propto S(\vec{Q},E) \\
S(\vec{Q},E)\propto [n(E)+1] \chi''(\vec{Q},E) \nonumber
\label{mod_lor}
\end{eqnarray}

\noindent In the main text we modeled the broad quasielastic scattering using a modified Lorentzian lineshape that is characterized by a single energy scale defined by the linewidth $\gamma$:

\begin{eqnarray}
\chi''(E)=\chi_{0} {E \over {1+\left({E\over \gamma}\right)^{2}}}.
\label{mod_lor}
\end{eqnarray}

\noindent Here $\chi_{0}$ is related to the real part of the susceptibility.  Alternatively, one can model the scattering in terms of a damped harmonic oscillator (DHO).  This approach adds an extra parameter to the fit: the energy position $\hbar \Omega_{0}$.  The DHO lineshape has the following form:

\begin{figure}[t]
\includegraphics[width=9.45cm] {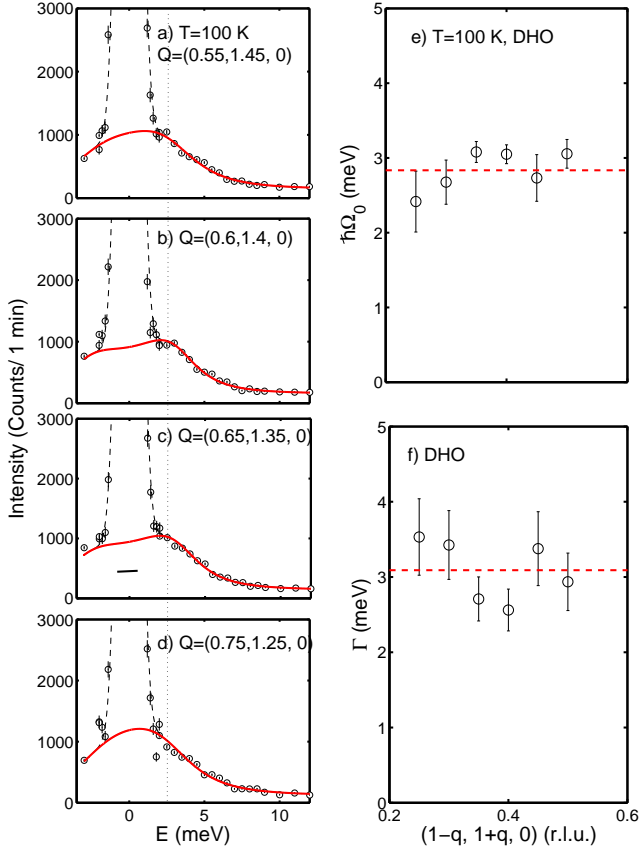}
\caption{\label{dso_q} The momentum dependence of the fluctuations near the $M$-point $\vec{Q}=(0.5,1.5,0)$ fit with the damped harmonic oscillator (Eqn.~\ref{DSO}) lineshape described in the appendix.  Panels $a-d)$ show constant-$\vec{Q}$ scans at various momentum transfers.  Panel $e)$ illustrates the momentum dependence of the fitted energy position and $f)$ shows the linewidth.  The results show that $\Gamma \sim \hbar \Omega_{0}$, with both parameters exhibiting little momentum dependence throughout the Brillouin zone.}
\end{figure}

\begin{eqnarray}
\chi''(E)= X_{0} \left({1 \over {1+\left({{E-\hbar\Omega_{0}} \over \Gamma} \right)^{2}}}-{1 \over {1+\left({{E+\hbar\Omega_{0}} \over \Gamma} \right)^{2}}} \right).
\label{DSO}
\end{eqnarray}

\noindent Here we denote the half-width by $\Gamma$ to distinguish it from that used in the modified Lorentzian lineshape in Eqn.~\ref{mod_lor}.  A key requirement of all neutron cross sections is that $\chi''$ be an odd function of energy; a property that both Eqn.~\ref{mod_lor} and \ref{DSO} possess.   In this Appendix we assess the results of the DHO analysis based on Eqn.~\ref{DSO} in terms of momentum transfer and temperature.

\begin{figure}[t]
\includegraphics[width=5cm] {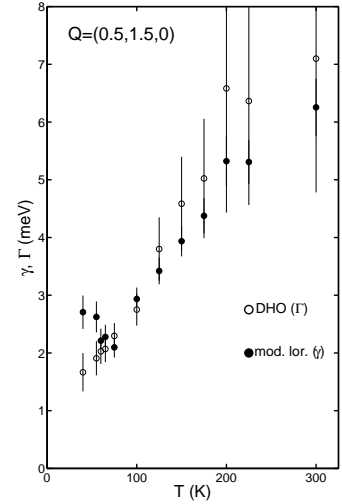}
\caption{\label{compare_param} A comparison of the temperature dependences of the linewidths $\Gamma$ of the DHO lineshape (Eqn.~\ref{DSO}) and $\gamma$ of the modified Lorentzian lineshape (Eqn.~\ref{mod_lor}).}
\end{figure}

Fig.~\ref{compare_lineshape} shows data from a constant-$\vec{Q}=(0.5,1.5,0)$ scan measured at 100\,K.  These data are fit to the modified Lorentzian lineshape (Eqn.~\ref{mod_lor}) in panel $a)$ and to the DHO lineshape (Eqn.~\ref{DSO}) ni panel $b)$.  It can be seen that while Eqn.~\ref{mod_lor} gives a good description of the data, Eqn.~\ref{DSO} fits the data slightly better because of the extra free parameter provided by the energy position $\hbar \Omega_{0}$.

We now investigate the momentum dependence of these parameters.  Fig.~\ref{dso_q} illustrates a series of constant-$\vec{Q}$ scans measured near the $M$-point $\vec{Q}=(0.5,1.5,0)$.   Panels $a-d)$ show constant-$\vec{Q}$ scans measured along $[1 \overline{1} 0]$ that were fit to a DHO lineshape convolved with the spectrometer resolution function (linewidth given by the horizontal solid bar shown in panel $c$).  (We note that the anomalously low intensity point in panel $d)$ near $\sim 1$\,meV represents a change that is much sharper than the instrumental energy resolution; it is also not reproduced at neighboring positions in momentum.  Thus we attach no physical significance to this isolated data point.)  Panels $e)$ and $f)$ display the momentum dependence of the energy position ($\hbar \Omega_{0}$) and linewidth ($\Gamma$), respectively.  Within our experimental uncertainties, both parameters are independent of $q$ and furthermore $\hbar \Omega \sim \Gamma$.  The average energy position is indicated by the vertical dashed line shown in panels $a-d)$.  From this we conclude that, independent of model (modified Lorentzian or DHO), the excitations probed here are overdamped and do not display the characteristics of propagating modes, which would manifest themselves as sharp, underdamped peaks in energy (like those shown in Fig.~1) that disperse with momentum.

Finally, we compare the temperature dependence of $\Gamma$, derived from Eqn.~\ref{DSO}, to that for $\gamma$, derived from Eqn.~\ref{mod_lor}, from which an activation energy was extracted in the main text.  Fig.~\ref{compare_param} displays both $\gamma$ and $\Gamma$ as functions of temperature.  It can be seen that both parameters agree within error.  (The DHO fit yields larger error bars because of the additional free parameter.)  The value of the activation energy, which was discussed in the main text, is therefore independent of analysis.

In summary, in this Appendix we have shown that the quasielastic fluctuations we observe in KLT(0.02) do not correspond to dispersive, well-defined, and underdamped excitations.  We have also shown that the modified Lorentzian and DHO models yield a consistent answer for the activation energy.  Given the extra parameter of the DHO model and the ambiguity in interpreting the energy position parameter $\hbar \Omega_{0}$ due to its lack of $q$ within the Brillouin zone, we favor the description provided by the modified Lorentzian, which is characterized by a single energy scale.  This is the analysis presented in the main text of the paper.  We note that more complex lineshapes with multiple energy scales might also possibly describe our observations.  An example is described in the context of spin fluctuations in superconducting YBa$_{2}$Cu$_3$O$_{6.353}$~\cite{Stock06_2:73,Stock08:77} or SrTiO$_3$.~\cite{Shapiro72:6}   But given the lack of an underdamped peak in our neutron data, we have not found this analysis to produce underdetermined parameters.


%

\end{document}